\documentclass[acmengage,9pt]{acmart}

\AtBeginDocument{%
  \providecommand\BibTeX{{%
    \normalfont B\kern-0.5em{\scshape i\kern-0.25em b}\kern-0.8em\TeX}}}

\setcopyright{cc}
\setcctype[4.0]{by}
\copyrightyear{2023}
\acmYear{2023}
\acmConference[HPCASIAWORKSHOP 2023]{International Conference on High Performance Computing in Asia-Pacific Region Workshops}{February 27-March 2, 2023}{Raffles Blvd, Singapore}
\acmBooktitle{\textup{This is the authors's version of the work.}
International Conference on High Performance Computing in Asia-Pacific Region Workshops (HPCASIAWORKSHOP 2023), February 27-March 2, 2023, Raffles Blvd, Singapore.}
\acmDOI{10.1145/3581576.3581610}
\acmISBN{978-1-4503-9989-0/23/02}
\usepackage{algorithmic}

\begin{document}

\title{Wilson matrix kernel for lattice QCD on A64FX architecture}

\author{Issaku Kanamori}
\authornote{All authors contributed equally to this research.}
\orcid{0000-0003-4467-1052}
\authornotemark[1]
\affiliation{%
  \institution{RIKEN Center for Computational Science}
  \streetaddress{7-1-26 Minatojima-minami-machi, Chuo-ku}
  \city{Kobe}
  \country{Japan}
  \postcode{650-0047}
}
\email{kanamori-i@riken.jp}

\author{Keigo Nitadori}
\orcid{0000-0001-7374-4236}
\affiliation{%
  \institution{RIKEN Center for Computational Science}
  \streetaddress{7-1-26, Minatojima Minamimachi, Chuo-ku}
  \city{Kobe}
  \country{Japan}
  \postcode{650-0047}
  }
\email{keigo@riken.jp}

\author{Hideo Matsufuru}
\orcid{0000-0003-1056-3969}
\affiliation{%
  \institution{Computing Research Center,
  High Energy Accelerator Research Organization (KEK) and
  School of High Energy Accelerator Science, Graduate University
  of Advanced Studies (SOKENDAI)}
  \streetaddress{1-1 Oho}
  \city{Tsukuba}
  \country{Japan}
  \postcode{305-0801}
}
\email{hideo.matsufuru@kek.jp}

\begin{abstract}
We study the implementation of the even-odd Wilson fermion matrix
for lattice QCD simulations on the A64FX architecture.
Efficient coding of the stencil operation is investigated for
two-dimensional packing to SIMD vectors.
We measure the sustained performance on the supercomputer Fugaku at
RIKEN R-CCS and show the profiler result of our code, which may signal
an unexpected source of slow-down in addition to the detailed efficiency
of each part of the code.
\end{abstract}



\maketitle

\section{Introduction}
Lattice QCD has been one of the typical benchmark applications in high performance computing,
in particular on massively parallel systems.
It has been also the case for the post-K (supercomputer Fugaku) project in Japan.
As one of the target applications of so-called Co-design project \cite{codesign}
in development of the A64FX architecture for Fugaku \cite{a64fx},
a high performance lattice QCD library has been developed and
released as the ``QCD Wide SIMD'' (QWS) library \cite{Ishikawa:2021iqw}.
The QWS library implements a linear equation solver, a typical bottleneck in lattice QCD simulations.
It is a solver for the clover fermion matrix, which is one of the popular fermion operators.
While QWS library achieves a good sustained performance of more than 100 PFlops, it has several restrictions:
it implements only the clover fermion matrix, the site degree of freedom only in the $x$-direction is packed into a SIMD vector, 
and it requires specific parameter setups to achieve good performance.
There is another general purpose lattice QCD code set ``Bridge++''
\cite{Ueda:2014rya,Bridge}
for which an optimized code branch for A64FX \cite{Akahoshi:2021gvk} is being developed.
This branch of Bridge++, named ``QXS'' after QWS, was originally developed as an interface
to call the QWS library and thus the convention and data layout are the same as QWS.
In addition, the QXS branch has implementation of several popular fermion matrices other than
the clover fermion, while the above mentioned restrictions are relaxed in the QXS.
In particular, the site degree of freedom in $x$-$y$ plane is packed into SIMD
vectors, which is useful for practical simulations of lattice QCD.
This generalization makes the shifts of field in $x$- and $y$-directions, that are
necessary for the stencil calculation, more involved to retain sufficient performance.

In this paper, we focus on one of such fermion operators, the even-odd version of
the Wilson fermion matrix, as a representative example of matrix that requires
such stencil calculation.
While A64FX shares the features of SIMD architecture with the same 512-bit
length with the Intel AVX-512, different techniques are required to achieve
sufficient arithmetic performance.
It would be informative to summarize such tuning procedures for the stencil
calculation on the A64FX architecture.
We also address the experience of using a profiler to find out an unexpected source
of performance decrease.

This paper is organized as follows.
In the next section, we briefly summarize the target fermion matrix
and related works.
In Section~\ref{sec:Implementation},
we describe our implementation of the Wilson fermion matrix
on the A64FX architecture.
Section~\ref{sec:Performance_benchmark} exhibits the performance
measurement and profiling experience on the Fugaku system.
The last section is devoted to summary and outlook.

\section{Fermion matrix in lattice QCD}
\label{sec:Fermion_matrix}

Quantum Chromodynamics (QCD) is the fundamental theory of
the strong interaction among quarks and gluons. 
The lattice QCD is a field theory formulated on a Euclidean
space-time lattice.
The path integral quantization enables numerical simulations based
on the Monte Carlo algorithm.
In lattice QCD, the gauge field (gluons) is represented with
$3\times 3$ complex matrix field, $U_\mu(x) \in SU(3)$,
that is defined on links connecting site $x$ and neighboring
site $x+\hat{\mu}$ where $\hat{\mu}$ is a unit vector along
the $\mu$-th axis.
Since the fermion fields, representing the quarks, are expressed
as anti-commuting Grassmann numbers, they are integrated out
and the result is a determinant of the fermion matrix $D$, which depends on
the background gauge field.
The determinant is evaluated as%
\footnote{This expression is for two quarks,
as one can show $\det D^\dagger D = (\det D)^2$ using a symmetry of $D$.
In practical simulations, Schur decomposition of the determinant based on
the even-odd operator treated in this article is often used.}
$\det D^\dagger D = \exp[- |D^{-1}\phi|^2]$, where $\dagger$ denotes hermitian
conjugate and $\phi$ is a Gaussian noise vector.
This requires one to compute $D^{-1} \phi$ during the Monte Carlo steps many times.
In the standard Hybrid Monte Carlo algorithm widely used in lattice QCD simulations,
one Monte Carlo step requires $O(10)$--$O(100)$ solving, in which $D$ is different
for each solving as the background gauge field changes.
In addition to the Hybrid Monte Carlo steps, calculation of an expectation
value often requires $D^{-1} \phi'$, where the right-hand side vector
$\phi'$ depends on the target quantity.  
In this way,
solving linear equations of the form $D \psi =\phi$ is a major
bottleneck of lattice QCD simulations.
As $D$ is a large sparse matrix, iterative solver algorithms are applied
to solve the linear equations, whose performance depends on the performance
of multiplication of $D$.

Since the lattice QCD simulations are performed with
finite lattice spacing, the final results are obtained after
extrapolation to the continuum limit.
The requirement for the fermion and gauge actions on the lattice is
that they coincide with the continuum actions in the continuum limit, 
and thus their discretized forms including the fermion matrix $D$ are not unique.
For the fermion action, several forms are used
together with various improvements.
In this paper, we focus on the Wilson fermion action,
of which fermion matrix is denoted as $D_{\mathrm{W}}$ in the following.  
It is one of the most popular fermion actions and its improved version
is the clover fermion implemented in QWS.

\begin{figure}[tbp]
  \includegraphics[scale=0.4]{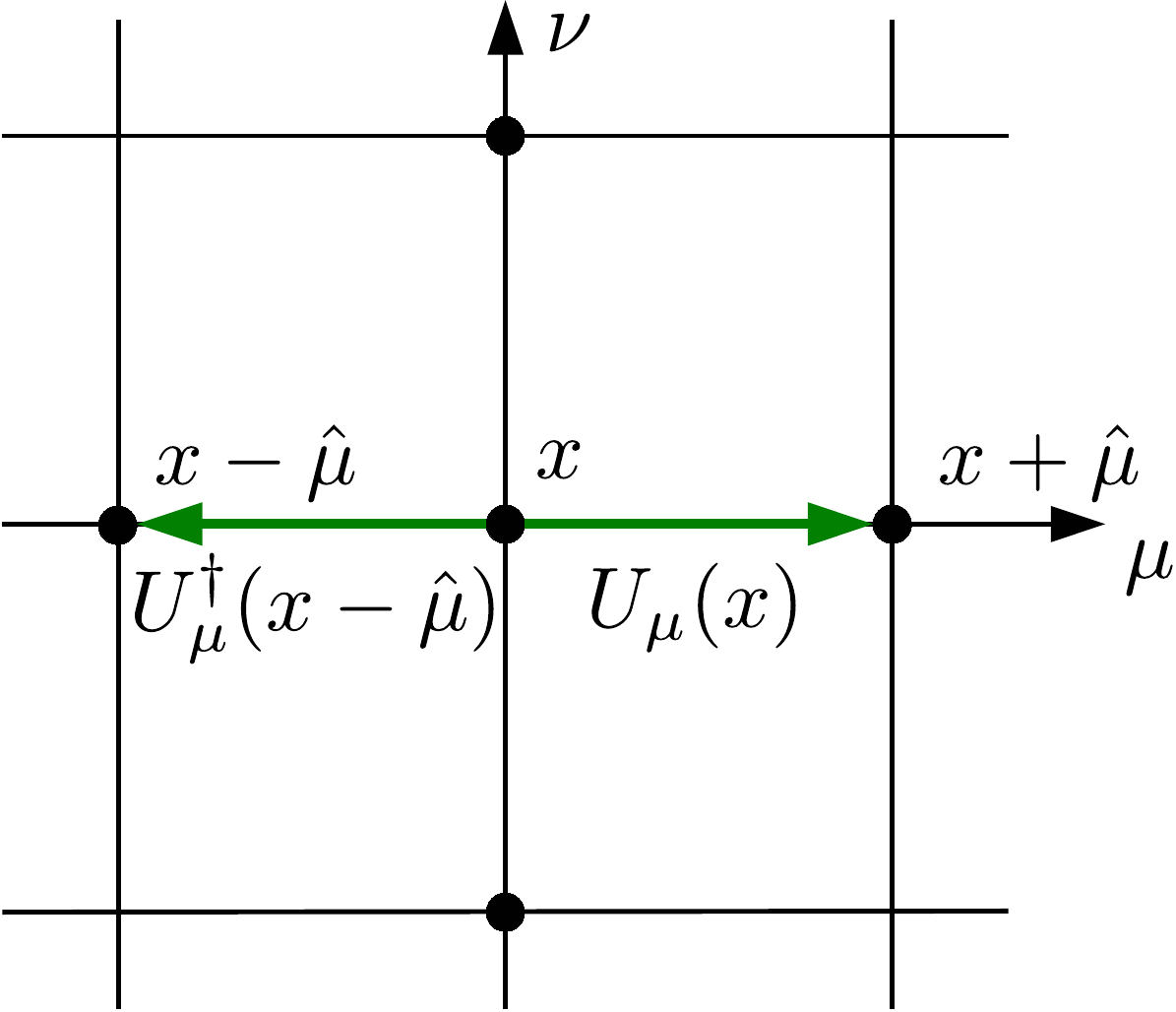}
  \caption{
  Schematic representation of the Wilson fermion matrix.
  }
  \label{fig:Wilson_matrix}
\end{figure}

The Wilson fermion matrix is written as\footnote{
By explicitly denoting the labels of matrix elements of $\gamma_\mu$
and $U_\mu$, the Wilson fermion matrix is
\begin{align*}
&(D_{\mathrm{W}})_{i a, j b}(x,y)
= \delta_{i,j}\delta_{a,b}\delta_{x,y}\\
& {}- \kappa \sum_{\mu=1}^4 \left[
  (1-\gamma_\mu)_{ij} (U_\mu)_{ab}(x)\delta_{x+\hat{\mu},y} +
  (1+\gamma_\mu)_{ij} (U_\mu)_{ba}^*(x-\hat{\mu})\delta_{x-\hat{\mu},y}
\right].
\end{align*}
The component of a vector on a lattice site $x$ is
$\phi_{ia}(x)$ ($i=1,\dots, 4$ and $a=1,2,3$).
}
\begin{align}
&D_{\mathrm{W}}(x,y) \nonumber\\
&= \delta_{x,y} - \kappa \sum_{\mu=1}^4 \left[
  (1-\gamma_\mu) U_\mu(x)\delta_{x+\hat{\mu},y} +
  (1+\gamma_\mu) U_\mu^\dag(x-\hat{\mu})\delta_{x-\hat{\mu},y}
\right],
\label{eq:Wilson_matrix}
\end{align}
where $x$, $y$ are lattice sites, $\gamma_\mu$ is a $4\times 4$
matrix acting on the spinor degree of freedom, $\kappa$ is
a parameter related to the fermion mass $m$ as $\kappa=1/(8+2m)$.
$U^\dag_\mu$ is a Hermitian conjugate of $U_\mu$.
$D_\mathrm{W}(x,y)$ contains only the coupling between neighboring sites
as shown in Figure~\ref{fig:Wilson_matrix}.
Since the matrix $\gamma_\mu$ has only one nonzero element in
each row and its value is $\pm 1$ or $\pm i$, its operation
is essentially interchange of the components.
Moreover, $\gamma_\mu$ satisfies $(\gamma_\mu)^2=1$ so that the factor $(\gamma_\mu \pm 1)$ works as
(two times) projection matrices.
The four components of spinor ($\phi_{ia}$ with $i=1,\dots, 4$)
is decomposed 
into two sets of two-component spinors ($\phi^{(\pm)}_{ia}$ with $i=1,2$), to which the
link variable $U_\mu(x)$ or $U_\mu^\dag(x-\hat{\mu})$ is multiplied to the color degrees of freedom.
After the link variable multiplication, one can easily reconstruct a four-component spinor.
For illustration purpose, an example of serial version of pseudo code is listed in Fig.~\ref{fig:Wilson_code}.
Note that at lines 5 and 8 in the figure the right hand side vector is shifted with respect to the lattice sites to generate the left hand side
for the stencil computation.
The number of floating point operations is 1368 FLOP per lattice site%
\footnote{
It actually depends on the details of $\gamma_\mu$ representation and the value
sited here is for the convention used in the QXS branch of Bridge++. 
}
and the B/F ratio is 1.12.

\begin{figure}
\begin{algorithmic}[1]
\FOR{$x$ in sites}
\STATE $\psi(x)=0$
\FOR{$\mu$ in $x$-, $y$-, $z$-, $t$- directions}
  \STATE $\phi^{(+)}(x+\hat{\mu}) ={}$ two-component spinor made from\\ $\allowbreak{(1+\gamma_\mu)\phi(x+\hat{\mu})} $
  \STATE multiply $3\times 3$ link variables: $\phi^{(+)}(x) := U_\mu(x) \phi^{(+)}(x+\hat{\mu})$
  \STATE reconstruct four-component spinor from\\ $\phi^{(+)}(x)$ and accumulate to $\psi(x)$
  \STATE $\phi^{(-)}(x-\hat{\mu}) = {}$
   two-component spinor made from\\ 
  $\allowbreak{(1-\gamma_\mu)\phi(x-\hat{\mu})} $
  \STATE multiply $3\times 3$ link variables:\\ \qquad ${\phi^{(-)}(x) := 
  U_\mu^\dagger(x-\hat{\mu}) \phi^{(-)}(x-\hat{\mu})}$
  \STATE reconstruct four-component spinor from\\ $\phi^{(-)}(x)$ and accumulate to $\psi(x)$
\ENDFOR
\ENDFOR
\end{algorithmic}
\caption{Pseudo code of a serial version of Wilson fermion matrix operation: $\psi = D_{\mathrm{W}} \phi$.}
\label{fig:Wilson_code}
\end{figure}

The equation to be solved in lattice QCD simulations is
\begin{equation}
 \sum_{k} D_{\mathrm{W}}(j,k) \xi (k) = \eta(j)
 \label{eq:linear_equation},
\end{equation}
where $\eta$ is a source vector, and
$j$ and $k$ are collective indices of site, color, and spinor
degrees of freedom.
The even-odd preconditioning is frequently adopted
to accelerate solving the above linear equation
(\ref{eq:linear_equation}).
Dividing the lattice sites into even and odd sites, 
Eq.~(\ref{eq:linear_equation}) is equivalent to 
\begin{equation}
    D_\mathrm{W}
    \begin{pmatrix}
      \xi _{\mathrm e} \\
      \xi _{\mathrm o}
    \end{pmatrix}
    =
    \begin{pmatrix}
      D_{\mathrm{ee}} & D_{\mathrm{eo}} \\
      D_{\mathrm{oe}} & D_{\mathrm{oo}} \\
    \end{pmatrix}
    \begin{pmatrix}
      \xi_{\mathrm e} \\
      \xi_{\mathrm o}
    \end{pmatrix}
    =
    \begin{pmatrix}
      \eta_{\mathrm e} \\
      \eta_{\mathrm o}
    \end{pmatrix}
    .
\end{equation}
Now $\xi_{\mathrm e}$ can be solved independently as in
\begin{equation}
 \left (1-D_{\mathrm{ee}}^{-1} D_{\mathrm{eo}} D_{\mathrm{oo}}^{-1} D_{\mathrm{oe}}\right) \xi_{\mathrm{e}}
 = D_{\mathrm{ee}}^{-1} \left(\eta_{\mathrm{e}} - D_{\mathrm{eo}} D_{\mathrm{oo}}^{-1} \eta_{\mathrm{o}}\right),
 \label{eq:even-odd_preconditioned_equation}
\end{equation}
and from that, $\xi_{\mathrm o}$ is obtained as
\begin{equation}
 \xi_{\mathrm{o}} = D_{\mathrm{oo}}^{-1} \left(\eta_{\mathrm{o}} - D_{\mathrm{oe}} \xi_{\mathrm{e}}\right).
\end{equation}
In the case of the Wilson fermion matrix, the diagonal
blocks $D_{\mathrm{ee}}$ and $D_{\mathrm{oo}}$ are just unit matrices
and their inverse is trivial.
Arithmetic operations of $D_{\mathrm{eo}}$ and $D_{\mathrm{oe}}$ are
in total the same as that of original $D_{\mathrm{W}}$.
However, the operator on the left hand side of
Eq.~(\ref{eq:even-odd_preconditioned_equation})
has, in most cases, a smaller condition number than the original
matrix $D_{\mathrm{W}}$, and thus it accelerates the solver algorithm.
In addition, the working vectors in iterative linear equation
solver requires half the size of full vectors, which has 
an advantage in memory access.

Here we briefly mention the related works on the tuning of lattice QCD
simulation codes on the A64FX architecture.
The situation just after the start of the shared use of 
supercomputer Fugaku was reviewed by Y.~Nakamura at the symposium on lattice field theory in 2021 (Lattice2021) \cite{nakamuraLat2021}.
The QWS library is a product of co-design project in the development
of Fugaku, and achieves sustained performance of more than 100 PFlops
\cite{Ishikawa:2021iqw}.
It is a linear equation solver for clover fermion matrix
and its implementation uses the Schwarz Alternating Procedure
which accelerates the iterative solver by domain-decomposed matrix
that reduces communication among the MPI processes.
To achieve the above high performance, the local lattice volume is chosen so that the data are on L2 cache.
The Bridge++ code set, the basis of this work, has developed various
fermion matrices \cite{Akahoshi:2021gvk}.
In particular, the five-dimensional formulation of the chirally
improved fermion action (domain-wall fermion) was also conducted
in the co-design project.
Combining the Bridge++ and QWS library, the multi-grid solver
for the clover fermion matrix is developed \cite{Ishikawa:2021day}.
One of the widely used lattice QCD libraries, Grid, has also implemented
the tuned code for A64FX architecture.
The result on QPACE 4 (an FX700 system) was reported in Lattice 2021
\cite{Meyer:2021uoj}.

\section{Implementation}
\label{sec:Implementation}

\subsection{SIMD features of A64FX architecture}
\label{subsec:SIMD_A64FX}

The A64FX architecture is based on the Armv8.2-A instruction sets
with the Scalable Vector Extension (SVE).
The A64FX processor consists of 48 compute cores and 2 or 4 assistant
cores, and 32 GiB on-package HBM memory.
These cores and memory are grouped into four Core Memory Groups (CMGs),
each having 12 compute cores sharing L2 cache of 8 MiB.
The peak performance depends on the operating frequency mode, 2.0 GHz
(normal mode) and 2.2 GHz (boost mode), and for the former
the peak performance of floating-point number operations is
3.072 and 6.144 TFlops for double and single precision, respectively.
The peak memory bandwidth per processor is 1024 GB/s.

Each node of Fugaku is composed of one A64FX processor, which is
connected to a Tofu interconnect D (TofuD) network.
TofuD network has a six-dimensional mesh-torus topology,
whose bandwidth amounts to 28 Gbps $\times$ 2 lanes $\times$ 10 ports.
In total, Fugaku has 158,976 nodes which correspond to 488 PFlops
of peak performance for double precision in the normal mode.

The A64FX architecture has SIMD registers and operation units
of 512-bit length.
While the SIMD length is the same as Intel AVX-512 instruction
set, prescription of optimization may differ significantly
(for our implementation and obtained performance on the latter,
see Ref.~\cite{Kanamori:2018hwh_inbook}).
One reason is that SIMD vector data types in Arm-SVE are sizeless,
and the SIMD length incorporated in the instruction set architecture
can take a range from 128- to 2048-bit.
We thus cannot declare an array of such a SIMD data type.

Manual manipulation of SIMD the instructions is possible
employing the Arm C Language Extension (ACLE) that is provided
through intrinsics.
Each intrinsic function directly specifies a SIMD operation
on SIMD variables.
Each vector instruction is accompanied by a predicate, that
masks the operation on the elements in the SIMD vectors.

Another difference in implementations of Arm-SVE and AVX-512
is in the treatment of complex numbers.
Compared to the latter, support of complex arithmetic
operations inside a SIMD vector is insufficient for Arm-SVE.
Such instructions, \textit{e.g.} \texttt{fcmla} that provides a mixed
operation for real and imaginary parts in a single SIMD vector,
were installed rather a late stage of development and are slow
due to its micro-code implementation \cite{a64fx} \footnote{
 Most simple floating-point instructions are executed on either pipeline
 A or B with a latency of 9 cycles, while most simple SIMD integer or
 shuffle instructions are executed on pipeline A only with a latency of 6 cycles
 \cite{a64fx}.
 The \texttt{fcmla} instruction is decomposed into two shuffle instructions
 and one floating-point instruction.
 Hence the throughput of this instruction is limited at most to one issue
 every two cycles.
}.

The instruction for floating point arithmetics is
rather straightforward.
Here we list the instructions used in the following
description for the shift of field in the form of
intrinsic functions.
For details of their operations are described in
\cite{Arm_ACLE}.
\begin{itemize}
\item LD1: basic load instruction to a SIMD vector register.
  In addition to the scalar base load instruction \texttt{svld1},
  there are ``gather-load'' instructions in which the indices of the memory address
  are specified by a supplemental integer vector.
\item ST1: basic store instruction from a SIMD vector to memory.
  Similarly to LD1, in addition to base store instruction
  \texttt{svst1}, there are ``scatter-store'' instructions
  in which the memory addresses are specified by an integer vector.
\item SEL: this instruction selects elements from two vectors according
  to a given predicate; from the first vector for active predicate bits and
  from the second for inactive bits.
\item TBL: arbitrary permutation of a given source vector according to
  an integer index vector.
  In a pseudo-code, it is like \texttt{dst[i] = src[idx[i]]}.
\item EXT: this instruction extracts consecutive elements of vector length
  from connected two vectors according to a given immediate integer value
  that specifies the head of the destination vector in the first source vector
  (\textit{cf.} Figure~\ref{fig:y-stencil}).
   There is a similar instruction ``SPLICE'' that accompanies a predicate.
   It extracts $n$ contiguous elements from one source vector
   according to the predicate, shifts another source vector for upward
   $n$ elements, and merges them.
\item COMPACT:
    this instruction collects active elements in a vector according to
    a predicate into lower consecutive elements of the vector and fills
    remaining elements with zero.

\end{itemize}

\subsection{Data layout}

\begin{figure}[tbp]
  \includegraphics[scale=0.6]{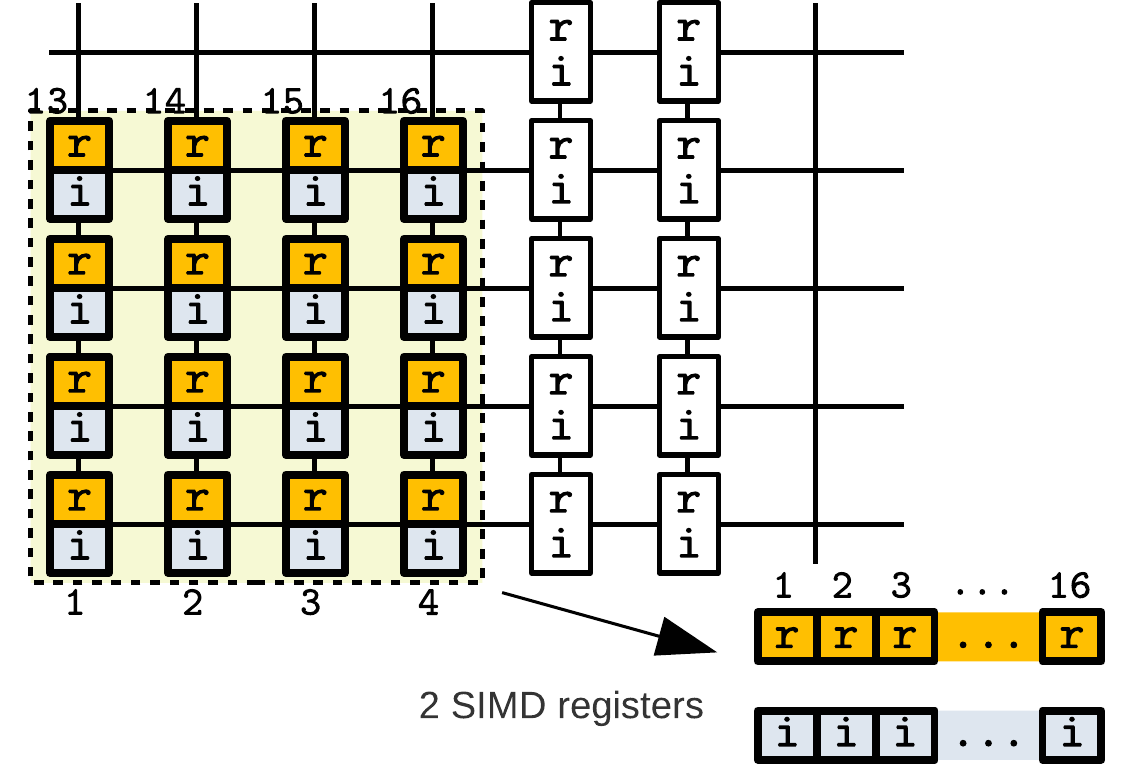}
  \caption{
  SIMD packing for the A64FX architecture in the case of $4\times 4$
  $x$-$y$ tiling for single precision.
  Real and imaginary parts of the complex numbers are stored in
  separate SIMD vectors.
  }
  \label{fig:SIMD_packing}
\end{figure}

The QWS library adopts packing the real and imaginary components
of a complex value into separate SIMD vectors, due to
the feature of arithmetic operations on complex numbers explained above.
We follow the SIMD packing strategy of QWS and also adopt packing the site degree of freedom into SIMD vector.
However, QWS packs the sites only in the
$x$-direction into a single SIMD vector, that restricts
the local lattice extent in the $x$-direction to be a
multiple of the vector length,
\textit{i.e.} 8 in double and 16 in single precision cases.
Moreover, these numbers are doubled for the even-odd matrix.
This restriction is practically inconvenient.
In the QXS branch of Bridge++, we thus decided to
enable 2D SIMD packing in $x$-$y$ plane,
as displayed in Figure~\ref{fig:SIMD_packing}.
Except for this extension, the data layout of QXS follows
the implementation of QWS.

The above SIMD packing procedure leads to the following data
layout.
For the even-odd site index and single precision,
the structure of the gauge and
spinor fields read
\begin{align}
 &\text{\texttt{float spinor[NT][NZ][NY/VLENY][NX/NEO/VLENX]}}\nonumber\\
&\text{\texttt{\hspace*{8em}[NC][ND][2][VLEN];}} \\
&\text{\texttt{float gauge[NDIM][NEO][NT][NZ][NY/VLENY]}}\nonumber\\
&\text{\texttt{\hspace*{6em}[NX/NEO/VLENX][NC][NC][2][VLEN];}}
\end{align}
where 
$\text{\texttt{NC}}=3$ for colors,
$\text{\texttt{ND}}=4$ for Dirac spinors, 
$\text{\texttt{NDIM}}=4$ for spacetime dimension,
$\text{\texttt{NEO}}=2$ for even-odd,
\texttt{NX}, \texttt{NY}, \texttt{NZ}, \texttt{NT} are the local lattice sizes
in $x$, $y$, $z$, $t$-directions,
\texttt{[2]} for real and imaginary components,
$\text{\texttt{VLEN}}=16$ (SIMD vector length),
and \texttt{VLENX} and \texttt{VLENY} are chosen so that 
$\text{\texttt{VLEN}}=\text{\texttt{VLENX}} \times \text{\texttt{VLENY}}$ ($\text{\texttt{VLENX}}\ge 2$).
These values are set as preprocessor macro constants except for
the local lattice sizes.
This data layout is a straightforward extension of the layout
of QWS to the $x$-$y$ SIMD tiling.
Regarding the complex, color, and spinor degrees of freedom as
a structure, the above layouts correspond to so-called
``Array of Structure of array (AoSoA)''.

Storing the real and complex parts of complex numbers to separate SIMD registers
makes a contrast to our implementation for 
Intel AVX-512 architecture~\cite{Kanamori:2018hwh_inbook}, in which the real
and complex parts are stored contiguously in a single vector.
Intel AVX-512 provides efficient complex operations for the elements
of a SIMD vector.
As a result, the number of sites packed into each vector is
halved so that a 1D data layout is adopted where \texttt{VLEN/2} sites in $x$-direction construct
each SIMD vector.

\subsection{Even-odd method}

\begin{figure*}[tbp]
  \includegraphics[scale=0.5]{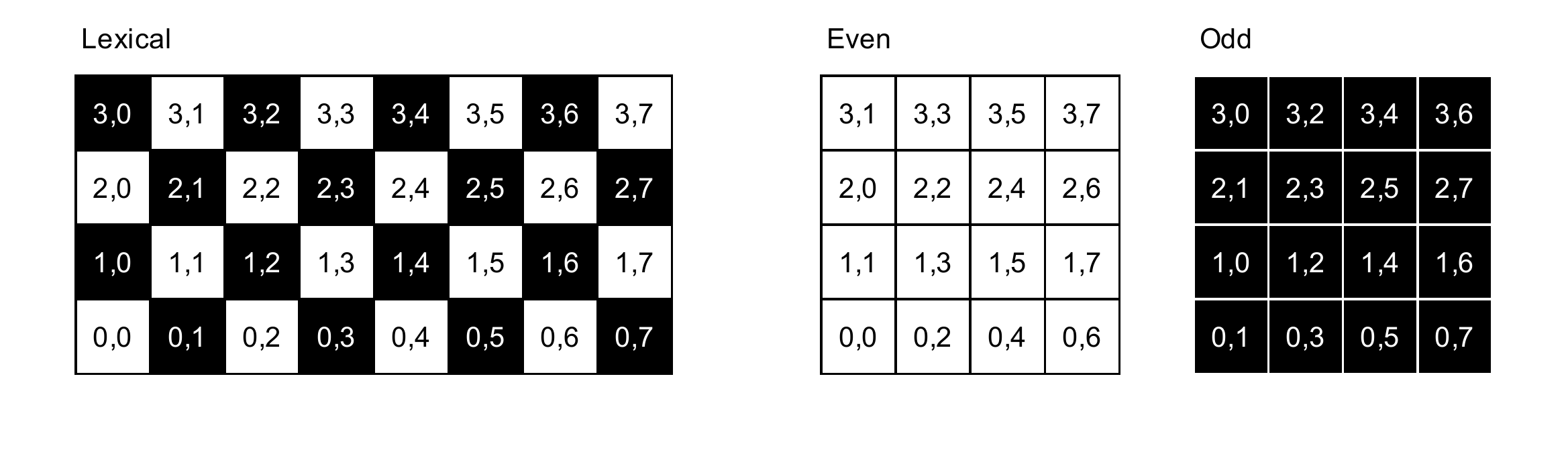}
  \caption{
  Left: Even and odd elements in an $x$-$y$ plane in a lexical form.
  White and black cells indicate even and odd elements, respectively,
  when the summation of the $z$ and $t$ indices is even.
  The numbers indicate the coordinates in $(y,x)$ form.
  Right: The even and odd elements are stored in separate arrays
  by compacting in the $x$-direction.
  }
  \label{fig:even-odd}
\end{figure*}

While the above $x$-$y$ SIMD packing largely relaxes the
restriction on the local volume, the shift of field in
$x$- and $y$-directions becomes somehow involved, in particular
for the even-odd version of fermion matrix.
Figure~\ref{fig:even-odd} explains the situation in the case
of $4\times 4$ SIMD packing for single precision.
The left panel shows the original lexical site index in the
$x$-$y$ plane with colors according to the site being even
(white) or odd (black) when the summation of the $z$ and $t$
indices is even.
The assigned numbers represent $(y,x)$ coordinate values.
The block matrices $D_{\mathrm{oe}}$ and $D_{\mathrm{eo}}$ act respectively on the
even and odd vectors whose components are stored in separate
arrays, as displayed in the right panel.
For example, $D_{\mathrm{oe}}$ is multiplied with an even vector and
results in an odd vector.
In more detail, the function representing $D_{\mathrm{oe}}$ reads
the data of an even vector, then apply eight points stencil
operations expressed in Eq.~(\ref{eq:Wilson_matrix}),
and write the result to an odd vector.
$D_{\mathrm{eo}}$ plays a similar role for odd and even vectors,
and for the Wilson fermion matrix $D_{\mathrm{ee}}$ and $D_{\mathrm{oo}}$
correspond to the multiplication of a unit matrix.
In the case of the clover fermion matrix, implemented in QWS
library as well as in  Bridge++, 
$D_{\mathrm{ee}}$ and $D_{\mathrm{oo}}$ are site-local block diagonal operations.

\subsection{Stencil access}

A stencil operation on a structured lattice, in general, requires
data on neighboring sites.
If the site is not in local but in another process, the data are transferred 
with message passing.
In our case, collecting the data on neighbor sites in $x$- and
$y$-directions requires merging of two SIMD vectors into one vector even in a bulk part.
One possible way to perform this operation
is to use the gather-load in loading the data from memory (see Subsection~\ref{subsec:SIMD_A64FX}).
 While the loading of data specifying the memory address by an index vector is convenient,
in practice, the gather-load (and scatter-store) is rather slow.
An alternative procedure is to load the data in units of SIMD vector and
rearrange them on registers exploiting the SVE instructions.
In what follows we describe the latter procedure for the even-odd
Wilson matrix.

\begin{figure*}[tbp]
  \includegraphics[scale=0.5]{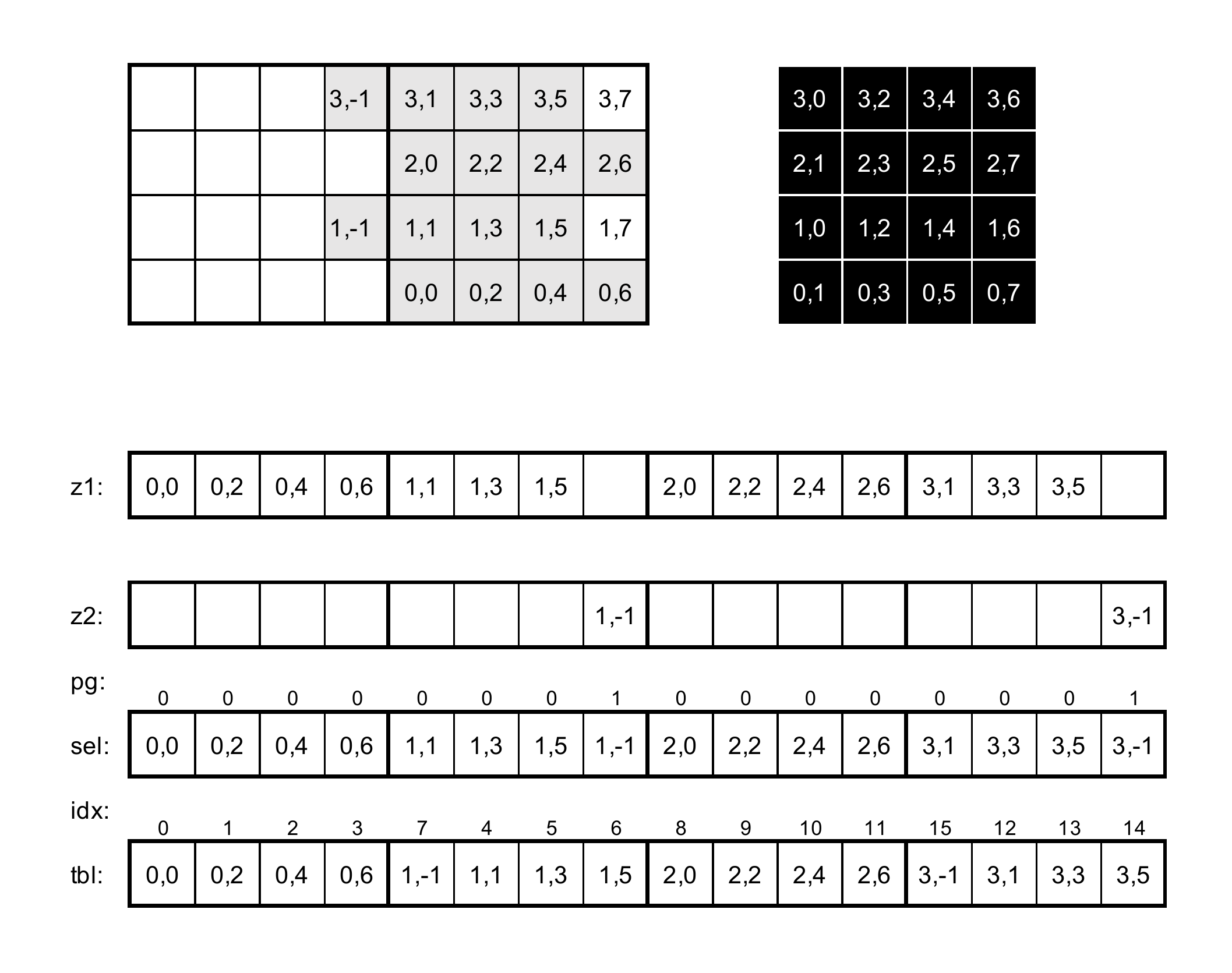}
  \caption{
  Shift of a field from the even array (top left) to the odd array (top right)
  in the stencil access of $D_{\mathrm{oe}}$.
  A $4 \times 4$ square region corresponds to a SIMD register or an aligned memory space.
  An $X^-$ operation requires access to the even arrays in irregular shape
  as indicated by light gray cells.
  The contents of the SIMD register are shown in the lower part of the figure
  (ordered from left to right).
  \texttt{z1}, \texttt{z2}:
  Results of load instructions from the current and left neighbor of the even array.
  \texttt{sel}: Result of a \texttt{sel} instruction of \texttt{z1} and \texttt{z2} with a predicate register \texttt{pg}.
  \texttt{tbl}: Results of a 
  \texttt{tbl}
  instruction which shuffles the previous result according to
  integer indices in another SIMD register \texttt{idx}.
  }
  \label{fig:x-stencil}
\end{figure*}

Since $D_{\mathrm{eo}}$ and $D_{\mathrm{oe}}$ contain the nearest
neighbor couplings, one needs to shift the field in 
$\pm \mu$-direction before multiplying $U_\mu(x)$ or after
multiplying $U_\mu^\dag(x-\hat{\mu})$.
Let us explain in more detail exemplifying $X^-$ contribution
(the second term in the square brackets in Eq.~(\ref{eq:Wilson_matrix}))
in $D_{\mathrm{oe}}$ in Figure~\ref{fig:x-stencil}.
Here we assume $4\times 4$ SIMD tiling in the single precision case
as an example, while we can choose other values of \texttt{VLENX}
and \texttt{VLENY}.
To shift the field in $+x$-direction, components from two SIMD vectors
are necessary to be loaded to construct a single SIMD vector.
According to the parity of $(y+z+t)$ coordinates, a part of the SIMD
vector is shifted as in the left panel of Figure~\ref{fig:x-stencil}.
For $D_{\mathrm{oe}}$, this operation acts on an even
vector (left) to generate an odd vector (right).
An $X^-$ operation requires access to the even arrays in irregular shape
as indicated by light gray cells.
We first load the current and left neighbor regions
to two registers \texttt{z1} and \texttt{z2}, respectively,
and use the \texttt{sel} instruction to merge them.
Then the data in the SIMD vector are rearranged with
\texttt{tbl} instruction.

\begin{figure*}[tbp]
  \includegraphics[scale=0.5]{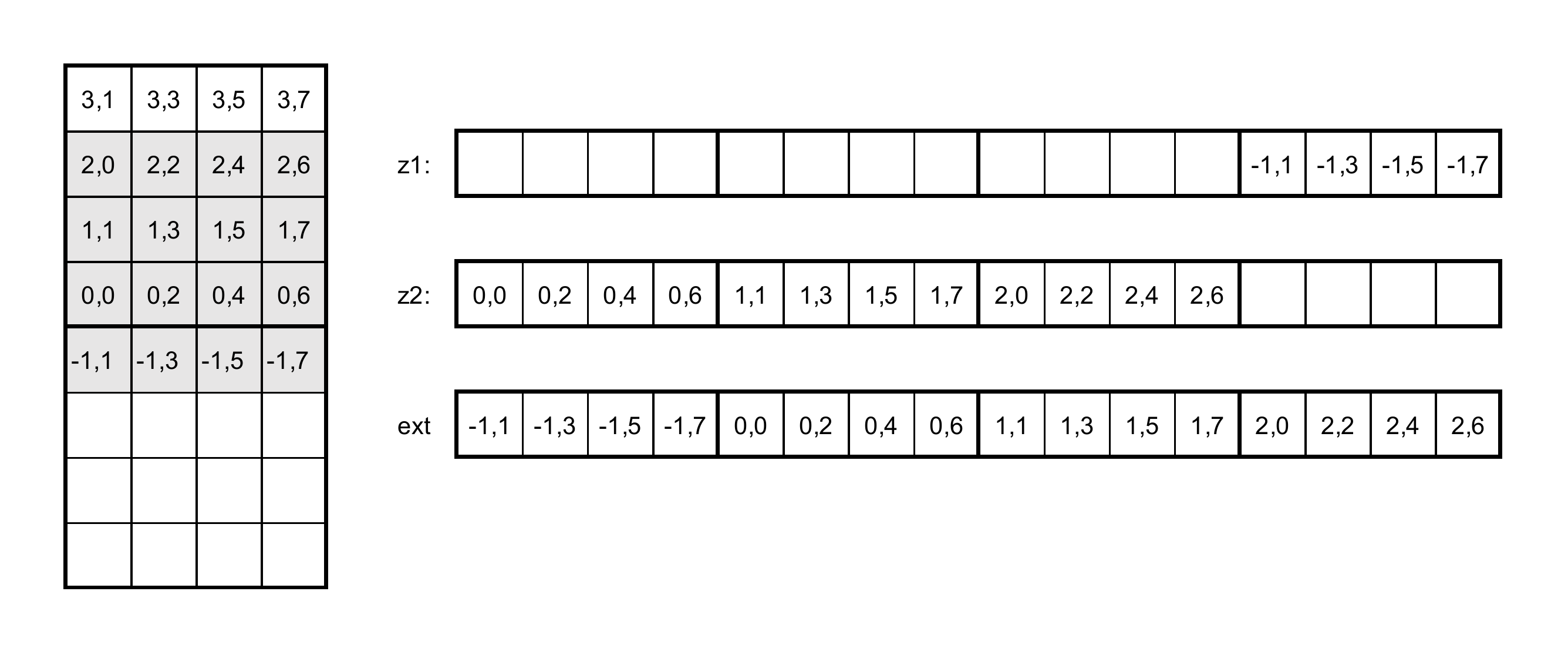}
  \caption{
  Similar to Fig.~\ref{fig:x-stencil} but shift for $Y^-$ operation.
  An \texttt{ext} instruction with an immediate value 12
  concatenates \texttt{z1} and \texttt{z2} and returns a shifted result. 
  }
  \label{fig:y-stencil}
\end{figure*}

The stencil access in $y$-direction is simpler than that in
$x$-direction, as displayed in Figure~\ref{fig:y-stencil}
for $Y^-$ operation.
We employ the \texttt{ext} instruction that extracts contiguous
data from two SIMD vectors and stores in a single vector.

\subsection{Communication}

\begin{figure*}[tbp]
  \includegraphics[scale=0.5]{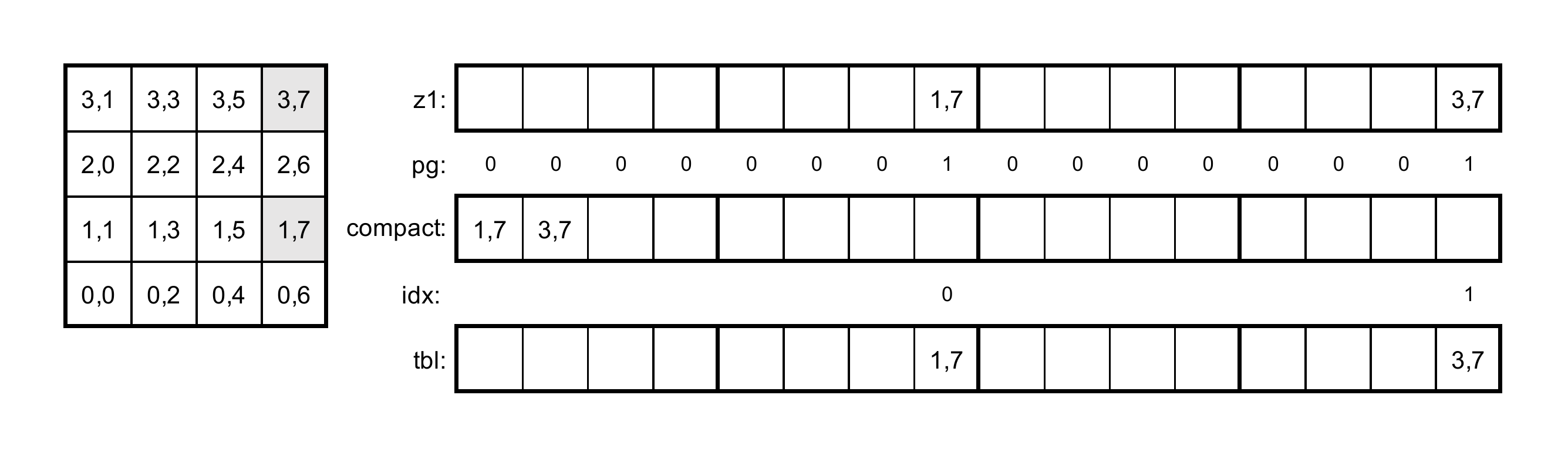}
  \caption{
  Packing and unpacking of data for MPI communications in the $X^-$ operation.
  Two of the sixteen elements in the SIMD register need to be sent.
  A \texttt{compact} instruction with a predicate indicating active elements
  collects them to the lower consecutive SIMD elements.
  After communication, the boundary data in the receive buffer are loaded to
  a SIMD vector and delivered to desired places
  by using the \texttt{tbl} instruction.
  }
  \label{fig:commbuf}
\end{figure*}

Since our code is parallelized with MPI, shifts of the SIMD elements
in the $x$- and $y$-directions require communication with neighboring processes.
For the communication, the boundary data are packed into
the buffer array, sent to the neighbor process, and unpacked to
merge with the data in the bulk region.
The communications and the operations in bulk region are overlapped.
Again communication in $x$-direction is involved as displayed in
Figure~\ref{fig:commbuf}.
We first implemented these packing and unpacking processes with
scatter-save and gather-load instructions with index vectors.
However, for the aforementioned reason, we replace these steps
with an alternative procedure as follows, which slightly increases
the performance.
After loading the boundary data into a SIMD vector,
the \texttt{compact} instruction is used to compress them
into lower elements of a temporal vector,
that are stored into the buffer array.
After the communication, the boundary data in the receive buffer are loaded to
a SIMD vector and delivered to desired places
by using the \texttt{tbl} instruction.
The shift in $y$-direction is again simpler than $x$-direction
because the boundary data are consecutive in a SIMD vector
that are stored to buffer array using the standard store instruction
with a predicate.
Unpacking is implemented in a similar manner with a masked load instruction.

\subsection{Multi-threading}
\label{subsec:Multithreading}

In Fujitsu's MPI libraries,
the thread mode \texttt{MPI\_THREAD\_MULTIPLE} is not available,
and thus communication is performed only by
master thread under the \texttt{MPI\_THREAD\_FUNNELED} mode.
Assuming that the assistant cores can execute background
communication, the tasks are averagely assigned to all threads
in the bulk part.

For the communication, boundary data are packed and unpacked
before and after the bulk part.
The packing of each direction is processed individually.
The loop in each direction is over a three-dimensional boundary
and averagely parallelized to the threads.
This process is denoted ``EO1'' in this paper.
Thus the load of each thread is well-balanced.
On the other hand, for unpacking, denoted ``EO2'',
a single loop for all the local sites processes the unpacking
of all the buffer data, since the number of boundaries concerning
each site depends on the place of the site on local lattice.
This causes load imbalance as we observe in the next section.

\section{Performance benchmark}
\label{sec:Performance_benchmark}

Performance is measured on the supercomputer Fugaku
at RIKEN Center for Computational Science (R-CCS).
We use a Fujitsu C/C++ compiler in its ``clang mode''
and the version of language and run-time is 4.8.1 tcsds-1.2.36.
Unlike the ``trad mode'' adopted in QWS, clang mode supports several new features:
full support of C++17, OpenMP 4.5, fp16 datatype, and ACLE for SVE.
We set several compile-time and runtime options for optimal performance.
The default value of the inline threshold in the clang mode turns out
to be too small.
To ensure that all eight hopping operations in the kernel function are
expanded inline, we set 
\texttt{-Rpass-missed=inline -mllvm -inline-threshold=1000}
in addition to \texttt{-Kfast} as the compiler options.
In the clang mode, the fast hardware barrier functionality within
the 12 cores of a CMG is not enabled by default.
To enable this feature, we set the environment variable \texttt{FLIB\_BARRIER =HARD}
whose impact to the performance amounts to about 20\% at our smallest lattice size.

\subsection{Profiler results}

We measure cycle account information by using
Fujitsu advanced performance profiler (FAPP).
The target code is the kernels extracted from the version of Bridge++ QXS branch
used in the presentation at Lattice 2022 symposium \cite{Aoyama:2022_lattice}.

This profiling is performed on one node of A64FX processor running at 2.2 GHz.
The total lattice size is
$L_x \times L_y \times L_z \times L_t = 16 \times 16 \times 16 \times 16$,
and the footprint of the gauge and spinor dataset is 18 MiB and 6 MiB,
respectively.
We use a standard combination of 4 MPI processes per node and 12 OpenMP
threads per process.
Assigning four processes to $[1,1,2,2]$ in parallelization,
the local lattice size per process is $16 \times 16 \times 8 \times 8$.
Applying the $4 \times 4$ SIMD tiling, each even/odd array consists of
$2 \times 4 \times 8 \times 8$ site elements.
To emulate a situation of the weak-scaling test in the next subsection,
we enforce the communication in $x$- and $y$-directions
for the self MPI process.

Due to a small lattice size intending on-L2 computation,
the execution period of each kernel function is short as $\lesssim 100$ $\mu$s.
In this time scale, the overhead of APIs in the profiler becomes non-negligible.
To mitigate this, we pass an option \texttt{-Hmethod=fast}
for the measuring command \texttt{fapp -C}, which bypasses the operating system
in reading hardware counters.

\begin{figure*}[tbp]
    \centering
    \includegraphics[width=0.75\linewidth]{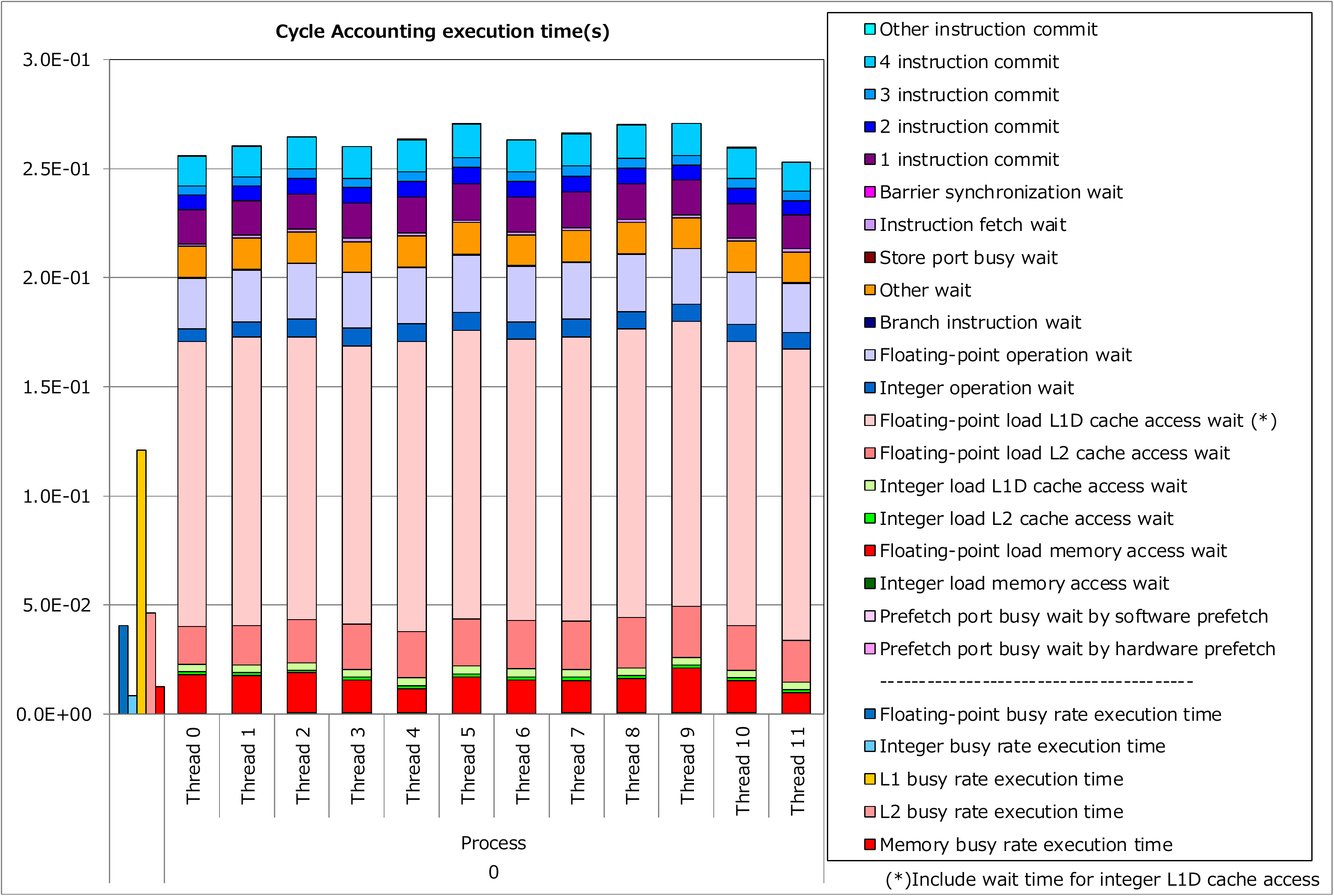}\\[1.0em]
    \includegraphics[width=0.75\linewidth]{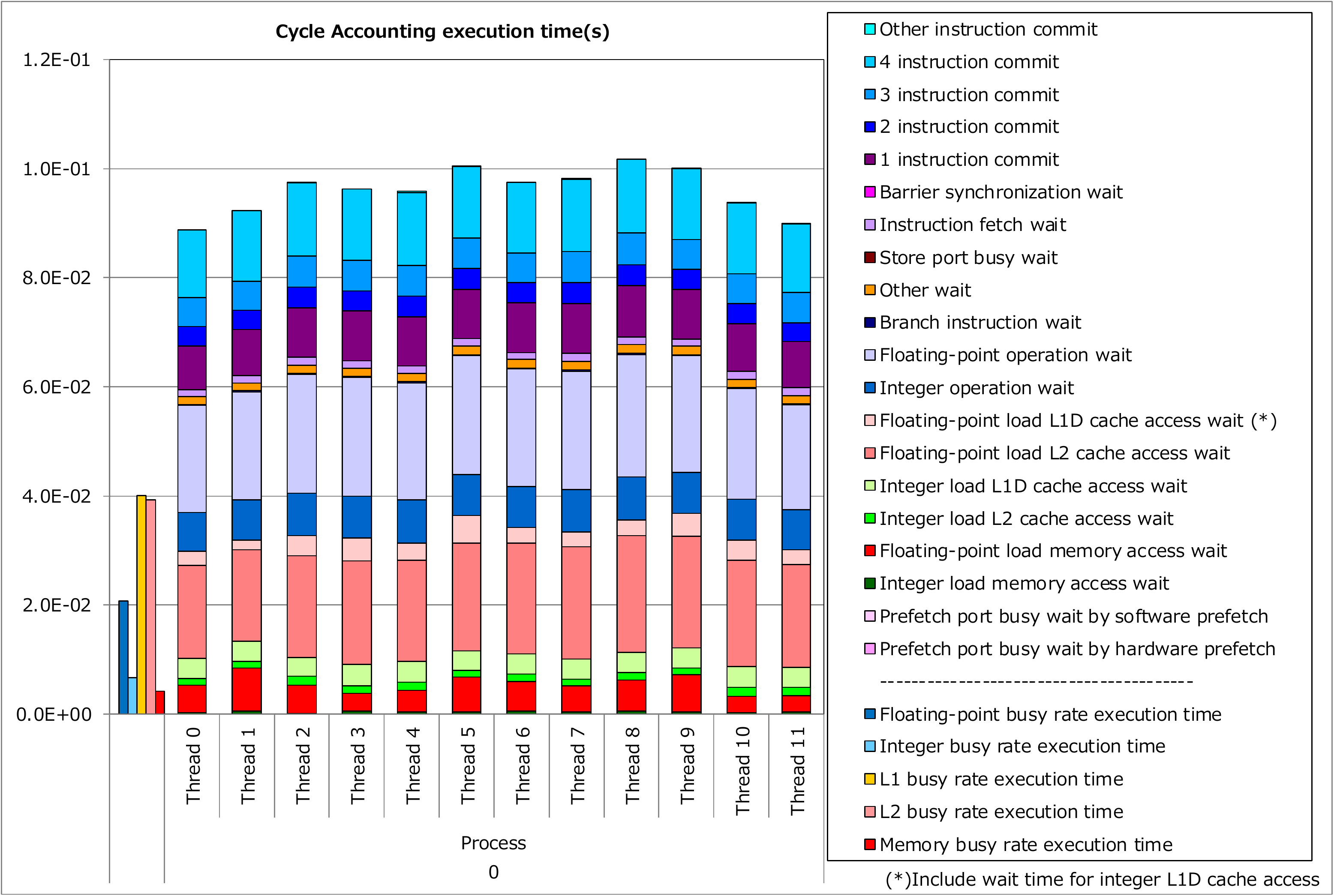}
  \caption{
  Cycle account information by FAPP:
  the initial result of the bulk part (top) and the result after tuning (bottom).
  The yellow bar on the leftmost column indicates the L1 cache busy time
  and pink colored parts in the twelve stacked bars are L1 cache wait time.
  The vertical axis shows the execution time in seconds after repeating
  the operation in Eq.~(\ref{eq:even-odd_preconditioned_equation}) 1000 times.
  Note that the scales of the vertical axis are different in the two panels.
  }
  \label{fig:prof-bulk}
\end{figure*}

\begin{figure*}[tbp]
  \centering
    \includegraphics[width=0.75\linewidth]{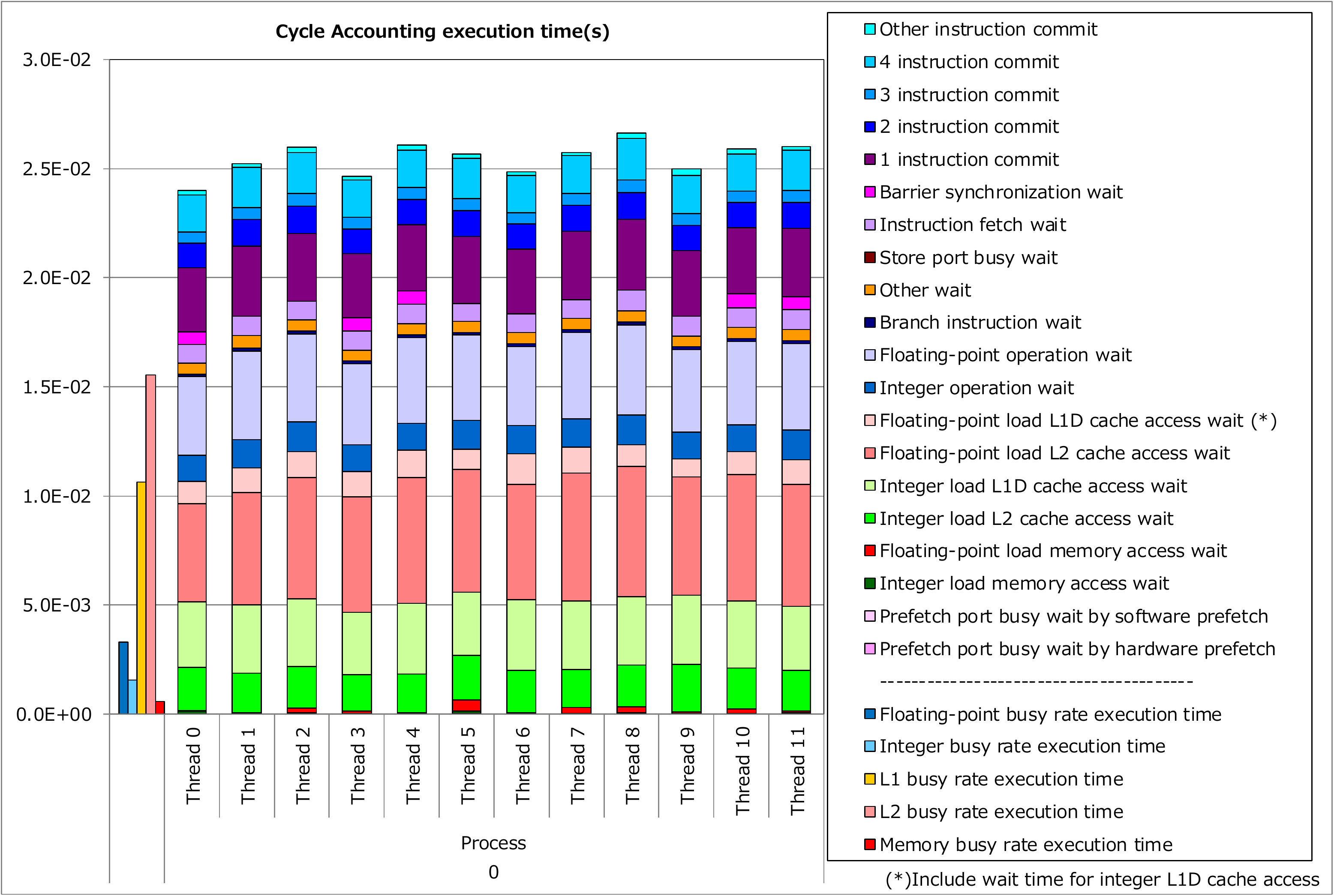}\\[1.0em]
    \includegraphics[width=0.75\linewidth]{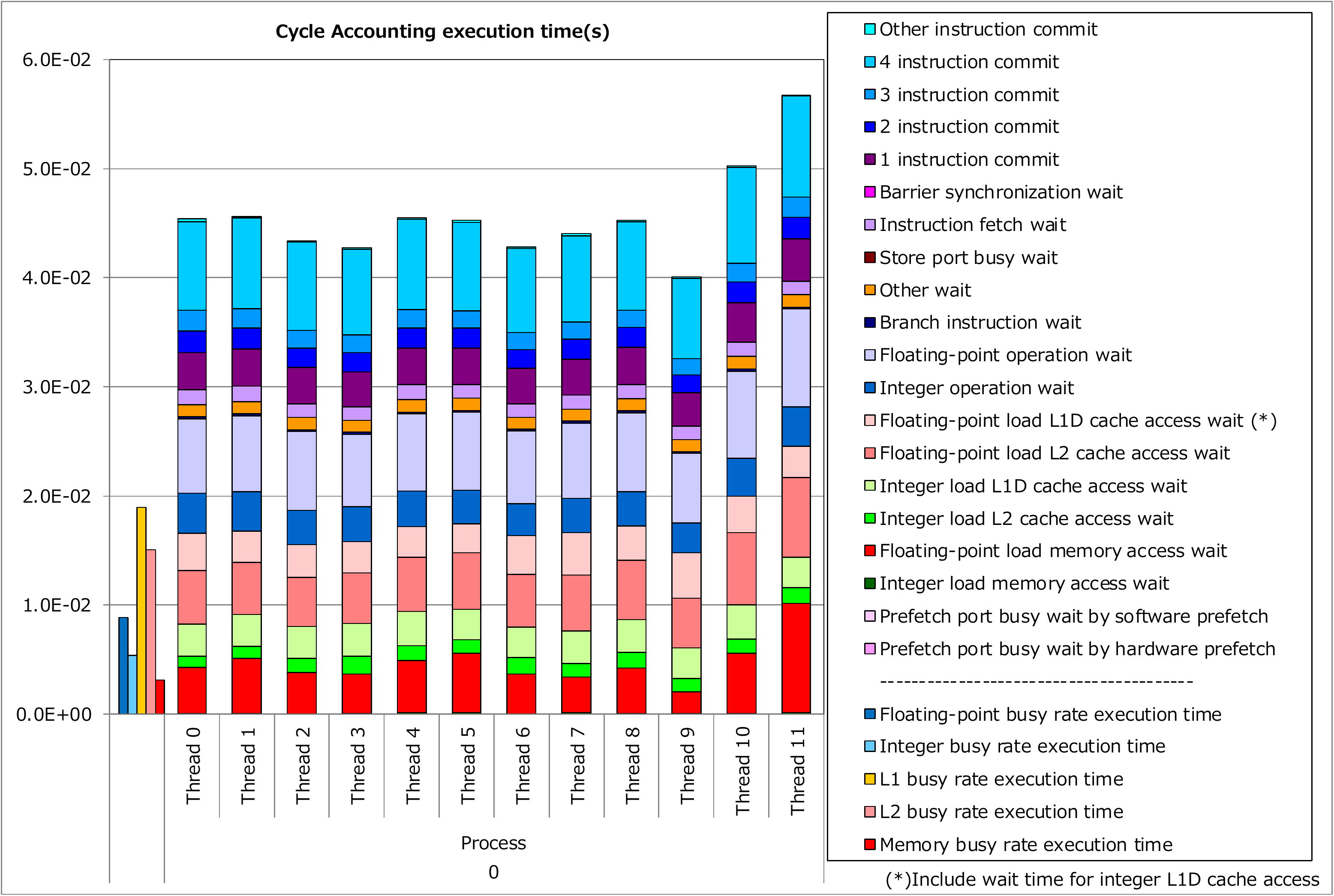}
  \caption{
  Cycle accounting information of the EO1 part for preparing the send buffer (top)
  and the EO2 part (bottom) for post-processing the received data.
  }
  \label{fig:prof-eo12}
\end{figure*}

Figure~\ref{fig:prof-bulk} shows the results of profiling the bulk part
before and after the improvement.
In the former case, heavy duty is imposed on the L1 cache resulting in
a bottleneck of the entire kernel.
This is an unexpected behavior regarding the property of a stencil kernel
--- it is usually memory or L2 bandwidth bottlenecked ---.
Careful examination of the profiler report reveals that some
fraction of the load and store instructions was occupied by
the gather-load and scatter-store instructions, even though
no such ACLE intrinsic function is used in the code.
It turned out that inefficient code was generated for the addition
of accumulated result of the stencil operations to the destination
array, and can be easily removed.
This part was composed of nested two loops, inner for elements of a
SIMD vector and outer for 24 (Re/Im)-spin-color degrees of freedom.
Such structure is prepared for developing a portable prototype code
expecting auto-vectorization by compilers for the inner loop.
While optimization by ACLE was applied to replace them,
such a code has remained in some parts of the code.
Presumably, the compiler interchanged the outer and inner loops resulting in
expensive stride accesses and gather-load/scatter-store instructions.
Replacing this data structure with explicit SIMD intrinsics or collapsed
single loop solves this pathological behavior.
This fact implies that the clang mode of the Fujitsu compiler is not fully matured.

Figure~\ref{fig:prof-eo12} shows the profiling results of the EO1 kernel for preparing
the send buffer and the EO2 kernel for post-processing the received data.
Both the kernels contain multiplications of $3\times 3$ gauge field matrix
for the data exported in upward
directions and the data imported from the upward directions.
Together with the bulk part in Figure~\ref{fig:prof-bulk}, we find
imbalance of execution time among the 12 threads, which is in particular sizable in EO2.
This imbalance is also confirmed by the number of floating point operations
executed on each core reported by the profiler.
The reason for the imbalance is that the local lattice sites are uniformly distributed to 12 threads as mentioned in Section~\ref{subsec:Multithreading}.
When the lattice site is located on the boundary, some of the operations 
in the bulk part is reduced and moved to the EO1 and EO2 kernels.
The load imbalance is significant in the EO2 kernel, in particular,
in thread 11 because it  processes more lattice sites on the boundary
and some of them are involved in multiple directions.
Especially, many of these sites receive the data from the upward process,
which requires multiplication of $3\times 3$ gauge field matrix.
In principle, the number of operations on each boundary lattice site can be
statically evaluated in advance.
In the future version, we plan to improve the load balance of the EO2 kernel
based on this empirical information.

\subsection{Performance of matrix multiplication}

We measure the sustained performance of the even-odd Wilson matrix
multiplication in the single precision case
on the supercomputer Fugaku in normal mode (2.0 GHz).
Throughout the performance measurements, we assign one MPI process
to one CMG, \textit{i.e.}, four MPI processes are assigned to each node.
In the following, each value of performance is measured in a single execution
by multiplying the matrix to a vector 1000 times.
Before the measurement, 300 seconds of sleep is inserted to avoid
the possible effect of remaining I/O of the previous job caused by
the file system of Fugaku.

The results of the sustained performance measured on a single node for
lattice sizes $16\times 16\times 16\times 16$,
$64\times 16\times 16\times 8$, and $64\times 32\times 32\times 16$
are summarized in Table~\ref{fig:2D_tiling_effect}.
We vary the 2-dimensional shape of the SIMD tiling:
$16\times 1$, $8\times 2$, $4\times 4$, and $2\times 8$.
The MPI process assignment is $[1,1,2,2]$ in $(x,y,z,t)$ directions.
The local lattice sizes for each process are listed
in Table~\ref{fig:2D_tiling_effect}.
As a part of the weak scaling measurement below, 
we enforce the communication in all four directions even
when only one MPI process exists in that direction.
For the smallest lattice, the data size is less than
the L2 cache size, which explains its better performance 
than the larger lattices.
There is no clear tendency against the shape of tiling, and the difference
is not significant.
Thus one can choose flexibly the values of \texttt{VLENX} and \texttt{VLENY}
depending on the setup of lattice and node parameters.

\begin{figure}[tbp]
  \includegraphics[width=0.98\linewidth]{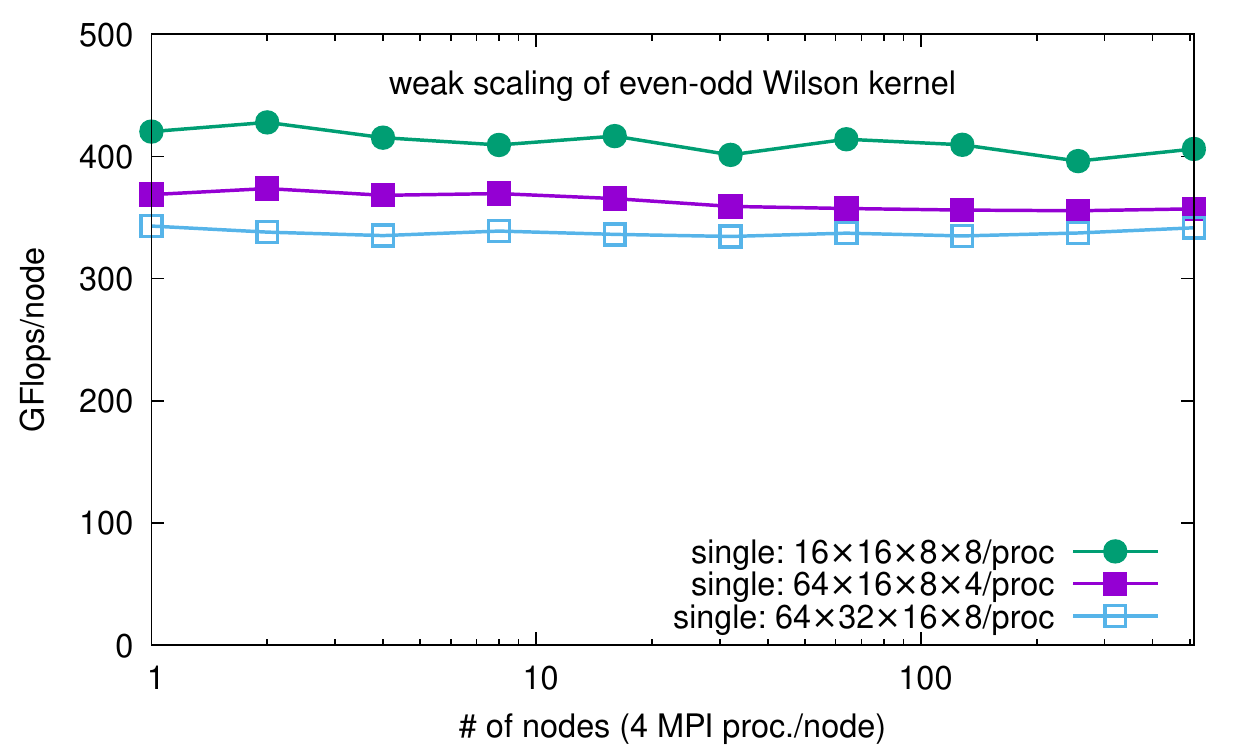}
  \caption{
  Weak scaling behavior of the sustained performance of the
  even-odd Wilson matrix multiplication in single precision
  on $16\times 16\times 8\times 8$, $64\times 16\times 8\times 4$,
  and $64\times 32\times 16\times 8$ local lattices per process.}
  \label{fig:weak_scaling}
\end{figure}

In Figure~\ref{fig:weak_scaling}, we show the weak scaling up to
512 nodes for the above three values of local lattice sizes,
$16\times 16\times 8\times 8$,
$64\times 16\times 8\times 4$, and $64\times 32\times 16\times 8$
with $4\times 4$ SIMD tiling.
The same enforcement of communication is applied in the measurements.
The rank maps are carefully prepared so that every neighboring communication
can be made within the same node (between CMGs) or with a neighboring node in the 6-dimensional mesh-torus network topology.
The performance per node is almost constant, implying very good
scaling up to 512 nodes we measured.
These local lattice sizes and numbers of nodes cover typical lattice sizes
in practical simulations in these years.

For comparison, we also measure the performance of our code
without ACLE implementation.
It is implemented in the same manner except for employing an array of
\texttt{float} of length \texttt{VLEN} instead of the builtin SIMD
data type, \texttt{svfloat32\_t}.
The ACLE intrinsics are replaced with inline functions that play
the same roles as the former.
On a single node of Fugaku, this results in around 30 GFlops,
about 10 times slower than the ACLE version,
which implies the vectorization by the compiler is not properly applied in this case.

\begin{table}[tbp]
  \caption{Effect of the 2D tiling for the even-odd Wilson fermion matrix
  multiplication in single precision.
  The performance is measured with four MPI processes on a single node.
  The lattice volume in the left-most column is per process.}
  \label{fig:2D_tiling_effect}
  \begin{tabular}{ccccccc}
    \toprule
    & \multicolumn{4}{c}{$x$-$y$ tiling} \\
   lattice size/process & $16\times 1$ & $8\times 2$ & $4\times 4$ & $2\times 8$ \\
    \midrule
    $16\times 16\times 8\times 8$   & $-$ & 448 & 420 & 419 \\
    $64\times 16\times 8\times 4$   & 339 & 343 & 369 & 380 \\
    $64\times 32\times 16\times 8$  & 319 & 328 & 343 & 345 &[GFlops] \\
    \bottomrule
  \end{tabular}
\end{table}

\section{Summary and outlook}

This paper concerns our implementation of the even-odd Wilson
fermion matrix for lattice QCD simulations on the A64FX architecture.
Efficient coding of the stencil operation is investigated for two-dimensional
packing to SIMD vectors.
While the shifts of SIMD data in $x$- and $y$-directions are involved,
in particular for the even-odd matrix, efficient implementation is
achieved by exploiting the ACLE intrinsics to shuffle the elements of
SIMD vectors instead of applying the gather-load and scatter-store instructions.
We measured the sustained performance on the supercomputer Fugaku at
RIKEN R-CCS as well as examined the result of profiling which sometimes
signals unexpected sources of slow-down.
The optimized code provides similar performance on other FX1000 systems,
while adjustment of setup, such as environment variables may be necessary.
These techniques are applicable to other fermion matrices in a straightforward
way and also to other stencil calculations in general.

The $x$-$y$ tiling of the SIMD vector has been applied throughout
the QXS branch of Bridge++.
Tuning developed in this paper is now being applied to the other fermion
matrices in the QXS branch.
The code developed for the A64FX architecture will be made publicly
available in the forthcoming release of Bridge++ version 2.0.

\begin{acks}
We thank Yoshifumi Nakamura and the members of Bridge++ project for many useful discussions.
For this work, discussions in HPC-physics workshop series were helpful.
This work is supported by JSPS KAKENHI (JP20K03961, JP22H01224),
the MEXT as ``Program for Promoting Researches on the Supercomputer Fugaku'' (Simulation for basic science: from fundamental laws of particles to
creation of nuclei, JPMXP1020200105) and ``Priority Issue 9 to be Tackled by Using
the Post-K Computer'' (Elucidation of The Fundamental Laws and Evolution
of the Universe),
and Joint Institute for Computational Fundamental Science (JICFuS).
The benchmarks and code tunings were performed on supercomputer Fugaku (through Usability Research ra000001) at RIKEN Center for Computational Science.
The code development was partially performed on the supercomputer
Flow at Information Technology Center, Nagoya University and Wisteria/BDEC-01 Odyssey
provided by Multidisciplinary Cooperative Research Program in Center for Computational Sciences, University of Tsukuba.
\end{acks}

\bibliographystyle{ACM-Reference-Format}
\bibliography{wilson_eo_kernel}

\end{document}